\documentclass[pra,aps,amsmath,amssymb,showkeys]{revtex4}
\usepackage[mathscr]{eucal}

\newcommand{\ch}{\mathscr{H}}
\newcommand{\xx}{\mathsf{X}}
\newcommand{\yy}{\mathsf{Y}}
\newcommand{\lsa}{{\cal{L}}_{s.a.}}
\newcommand{\piq}{\mathsf{\Pi}}
\newcommand{\qq}{\mathsf{Q}}
\newcommand{\rr}{\mathsf{R}}
\newcommand{\uu}{\mathsf{u}}
\newcommand{\vv}{\mathsf{v}}

\begin{document}
\clearpage
\preprint{}

\title{{A Shrinking Factor for Unitarily Invariant Norms under a Completely Positive Map}}
\author{Alexey E. Rastegin}
\affiliation{Department of Theoretical Physics, Irkutsk State University,
Gagarin Bv. 20, Irkutsk 664003, Russia}

\begin{abstract}
A relation between values of a unitarily invariant norm of
Hermitian operator before and after action of completely positive
map is studied. If the norm is jointly defined on both the input
and output Hilbert spaces, one defines a shrinking factor under
the restriction of given map to Hermitian operators. As it is
shown, for any unitarily invariant norm this shrinking factor is
not larger than the maximum of two values for the spectral norm and the
trace norm.
\end{abstract}


\maketitle

\pagenumbering{arabic}
\setcounter{page}{1}

\section{Introduction}

In many disciplines, linear maps on a space of
operators provide key tools for treatment of the subject. For
several reasons the class of completely positive maps (CP-maps) is
especially valuable \cite{bhatia07}. The recent advances in
quantum information theory have led to a renewed interest in this
area \cite{nielsen}. In effect, it seems that all possible changes
of quantum state is covered by contractive completely positive
maps \cite{kraus}, though the monotonicity of relative entropy can
be proved in more general framework \cite{uhlmann77}. Anyway, all
the important examples are actually completely positive. Thus,
studies of used quantitative measures under action of CP-maps form
a very actual issue. Of course, other properties of CP-maps with
respect to certain norms are subjects of active research
\cite{devetak,jenca}. As a rule, distance measures are metrics
induced by norms with  handy properties. Unitarily invariant norms
are very useful in this regard \cite{lidar1}. At the same time,
some variety of measures is typically needed with respect to
themes of interest. So, a question on contractivity of given map
with respect to applied norm is significant in many different
fields of physics (see \cite{ruskai} and references therein).
Hence we may be interested in general results on a problem of
contractivity without explicit specification of the measure. Below
the result of such a kind will be given for the class of unitarily
invariant norms. Namely, the norm of image of Hermitian operator
is not greater than the norm of operator itself multiplied by some
shrinking factor. For given CP-map and any norm from the
considered class, this factor does not exceed the maximum of two
exact values of shrinking factor for the spectral norm and the
trace norm. The discussion is carried out entirely in finite
dimensional setting.

\section{Definition and notation}

Let ${\ch}$ be $d$-dimensional Hilbert space. We
denote by ${\cal{L}}({\ch})$ the space of all linear operators on
${\ch}$, and by $\lsa({\ch})$ the space of self-adjoint
(Hermitian) operators on ${\ch}$. For any
${\xx}\in{\cal{L}}({\ch})$ the operator ${\xx}^{\dagger}{\xx}$ is
positive semidefinite, and its unique positive square root is
denoted by $|{\xx}|$. The eigenvalues of $|{\xx}|$ counted with
multiplicities are the singular values of operator ${\xx}$, in
signs $\sigma_i({\mathsf{X}})$ \cite{hornj}. Each unitarily
invariant norm is generated by some symmetric gauge function of
the singular values, i.e.
$|||{\xx}|||_{\rm{g}}={\rm{g}}\bigl(\sigma_1({\xx}),\ldots,\sigma_d({\xx})\bigr)$
(see, e.g., theorem 7.4.24 in \cite{hornj}). The determining
properties for a symmetric gauge function are listed in
\cite{hornj}. The two families, the Schatten norms and the Ky Fan
norms, are most widely used. For any real $p\geq1$, the Schatten
$p$-norm is defined as \cite{hornj}
\[
||{\xx}||_p:=\left(\sum\nolimits_{i=1}^{d} \sigma_i({\xx})^p \right)^{1/p}
\ .
\]
This family recovers the trace norm $||{\xx}||_{\rm{tr}}$ for
$p=1$, the Frobenius norm $||{\xx}||_F$ for $p=2$, and the
spectral norm $||{\xx}||_{\infty}$ for $p\to\infty$ \cite{hornj}.
Let us use these signs, although
$||{\xx}||_{\infty}\equiv||{\xx}||_{(1)}$ and
$||{\xx}||_{\rm{tr}}\equiv||{\xx}||_{(d)}$ as the Ky Fan norms
though. For integer $k\geq1$, the Ky Fan $k$-norm is defined by
\cite{hornj}
\begin{equation}
||{\xx}||_{(k)}
:=\sum\nolimits_{i=1}^{k} \sigma_i^{\downarrow}({\xx})\equiv
{\rm{g}}_{(k)}\bigl(\sigma_1({\xx}),\ldots,\sigma_d({\xx})\bigr)
\ , \label{kfndef}
\end{equation}
where the arrows down show that the singular values are put in the
decreasing order. In terms of the norms (\ref{kfndef}), the
partitioned trace distances have been introduced \cite{rast091}.
These measures enjoy similar properties to the trace norm
distance. In the following, we will assume that
$||{\xx}||_{(k)}\equiv||{\xx}||_{\rm{tr}}$ for $k\geq{d}$. We
shall now define the main object treated in this paper.

{\bf Definition 2.1.} {\it Let $\Phi_{s.a.}$ be the restriction of
CP-map $\Phi:{\cal{L}}({\ch}_A)\rightarrow{\cal{L}}({\ch}_B)$ to
Hermitian operators. Its shrinking factor with respect to given
unitarily invariant norm $|||\centerdot|||_{\rm{g}}$ is defined
as}
\[
\eta_{\rm{g}}(\Phi_{s.a.}):=\sup\Bigl\{|||\Phi({\xx})|||_{\rm{g}}:
{\>}{\xx}\in{\lsa}({\ch}_A),{\>}|||{\xx}|||_{\rm{g}}=1\Bigr\}
\ .
\]

If ${\ch}_A={\ch}_B$ then on both the spaces a norm
$|||\centerdot|||_{\rm{g}}$ is defined by the same symmetric gauge
function. When ${\rm{dim}}({\ch}_A)\neq{\rm{dim}}({\ch}_B)$, we
append zero singular values so that the vectors $\sigma({\xx})$
and $\sigma\bigl(\Phi({\xx})\bigr)$ have the same dimensionality
equal to $\max\{d_A,d_B\}$. In this regard, our consideration is
related to those symmetric gauge functions that are not changed by
adding zeros. Only under this condition the same unitarily
invariant norm is correctly defined on the spaces of different
dimensionality. The needed property is provided by all the
functions assigned to the Ky Fan norms and the Schatten norms. Any
linear combination of such functions with positive coefficients is
also a symmetric gauge function that enjoys this property. Indeed,
the symmetric gauge functions, providing the above property, form
a convex set.

In Definition 2.1 the supremum is taken over Hermitian inputs
${\xx}$. First, self-adjoint operators are very important in many
applications including quantum information topics. Say, the
difference between two density matrices is traceless Hermitian,
and the restriction to such operators deserves attention
\cite{ruskai}. Second, a consideration of Hermitian ${\xx}$ allows
to simplify analysis. Third, some relations with positive or
self-adjoint operators have later been extended to more general
ones \cite{bhatia87,watrous2}. So, our definition is suitable for
such a generalization.

In the seminal paper \cite{kyfan} Ky Fan obtained important
results with respect to extremal properties of eigenvalues. One of
his formulations is now known as Ky Fan's maximum principle. The
present author have applied this power principle for stating the
basic properties of the partial fidelities \cite{rast092}, which
were originally introduced by Uhlmann \cite{uhlmann00}, and the
partitioned trace distances \cite{rast091}. Changing the proof of
theorem 1 in \cite{kyfan}, we can merely prove
\begin{equation}
\sum\nolimits_{i=1}^{k} \lambda_i^{\downarrow}({\xx})=
\max\bigl\{{\,}{\rm{Tr}}({\mathsf{P}}{\xx}):
{\>}{\mathbf{0}}\leq{\mathsf{P}}\leq{\mathbf{I}},{\>}{\rm{Tr}}({\mathsf{P}})={k}\bigr\}
\ , \label{kfmaxpr02}
\end{equation}
where the maximum is taken over those positive operators
${\mathsf{P}}$ with trace $k$ that satisfy
${\mathsf{P}}\leq{\mathbf{I}}$. Alternately, the maximization may
be over all projectors of rank $k$, as in the original statement
\cite{kyfan}. If operator ${\xx}$ is positive semidefinite then
the maximum can be taken under the condition
${\rm{Tr}}({\mathsf{P}})\leq{k}$ or, for projectors,
${\rm{rank}}({\mathsf{P}})\leq{k}$. Using the Jordan
decomposition, we have the following result.

{\bf Lemma 2.2.} {\it For any ${\xx}\in\lsa({\ch})$ and $k\geq1$,
there exist two mutually orthogonal projectors ${\mathsf{P}}_Q$
and ${\mathsf{P}}_R$ such that
${\rm{rank}}({\mathsf{P}}_Q+{\mathsf{P}}_R)\leq{k}$ and}
\[
||{\xx}||_{(k)}={\rm{Tr}}\bigl[({\mathsf{P}}_Q-{\mathsf{P}}_R){\,}{\xx}\bigr]
\ .
\]

{\it Proof.} First, we suppose that $k\leq{d}$. We write
${\xx}={\qq}-{\rr}$ with positive semidefinite ${\qq}$ and ${\rr}$
whose supports are orthogonal. These operators are positive and
negative parts of ${\xx}$ respectively. Putting the spectral
decomposition
\[
|{\xx}|={\qq}+{\rr}=\sum\nolimits_q q{\,}{\uu}_q{\uu}_q^{\dagger}+
\sum\nolimits_r r{\,}{\vv}_r{\vv}_r^{\dagger}
\ ,
\]
we see that $\{q\}\cup\{r\}=\{\sigma_i({\xx})\}$. For given $k$,
we define two subspaces, namely
\[
{\mathscr{K}}_Q:={\rm{span}}\bigl\{{\uu}_q:{\>}q\in\{\sigma_1^{\downarrow},\sigma_2^{\downarrow},\ldots,\sigma_k^{\downarrow}\}\bigr\}
\ ,\quad
{\mathscr{K}}_R:={\rm{span}}\bigl\{{\vv}_r:{\>}r\in\{\sigma_1^{\downarrow},\sigma_2^{\downarrow},\ldots,\sigma_k^{\downarrow}\}\bigr\} \ .
\]
If ${\mathsf{P}}_Q$ is projector onto ${\mathscr{K}}_Q$ and
${\mathsf{P}}_R$ is projector onto ${\mathscr{K}}_R$, then we at
once get
$({\mathsf{P}}_Q-{\mathsf{P}}_R){\,}{\xx}={\mathsf{P}}|{\xx}|$ for
projector ${\mathsf{P}}={\mathsf{P}}_Q+{\mathsf{P}}_R$ of rank
$k$. By construction, the trace of ${\mathsf{P}}|{\xx}|$ sums just
$k$ largest singular values of ${\xx}$. The case $k>d$ is reduced
to the trace norm for that the needed projectors are already built
and ${\rm{rank}}({\mathsf{P}}_Q+{\mathsf{P}}_R)=d<{k}$. $\square$

\section{Main results}

In this section, we will study a change of unitarily
invariant norms under action of a CP-map. Since they are
positive-valued, upper bounds are usually indispensable. Let
$\Phi:{\cal{L}}({\ch}_A)\rightarrow{\cal{L}}({\ch}_B)$ be a
completely positive linear map. We shall use the Choi-Kraus
representation \cite{choi,kraus}
\[
\Phi({\xx})=\sum\nolimits_{n} {\mathsf{E}}_{n}
{\,}{\xx}{\,}{\mathsf{E}}_{n}^{\dagger}
\ , \quad {\mathsf{E}}_{n}:{\,}{\ch}_A\rightarrow{\ch}_B \ .
\]
From the physical viewpoint, this result is examined in
\cite{nielsen}. In the context of Stinespring's dilation theorem,
it is discussed in \cite{bhatia07}. The Choi-Kraus representation
is not unique, but a freedom is unitary in character (see theorem
8.2 in \cite{nielsen}). Two sets $\{{\mathsf{E}}_{n}\}$ and
$\{{\mathsf{G}}_{m}\}$ determine the same CP-map if and only if
\[
{\mathsf{G}}_{m}=\sum\nolimits_{n} v_{mn} {\mathsf{E}}_{n} \ ,
\]
where numbers $v_{mn}$ are entries of some unitary matrix of
proper dimensionality. Then for given CP-map the two positive
semidefinite operators
\[
{\mathsf{M}}:=\sum\nolimits_{n} {\mathsf{E}}_{n}{\mathsf{E}}_{n}^{\dagger}
\ , \quad {\mathsf{W}}:=\sum\nolimits_{n} {\mathsf{E}}_{n}^{\dagger}{\mathsf{E}}_{n}
\ ,
\]
are not dependent on a choice of the set $\{{\mathsf{E}}_{n}\}$.
The second operator has been used for another definition of the
trace norm distance via extremal properties of contractive CP-maps
\cite{rast07}.

{\bf Theorem 3.1.} {\it Let
$\Phi:{\cal{L}}({\ch}_A)\rightarrow{\cal{L}}({\ch}_B)$ be a
CP-map. For every ${\xx}\in{\lsa}({\ch}_A)$ there holds}
\[
||\Phi({\xx})||_{(k)}\leq \eta{\,}||{\xx}||_{(k)} \ ,
\quad k=1,2,\ldots,\max\{d_B,d_A\}
\ , 
\]
{\it where the factor
$\eta:=\max\left\{||{\mathsf{M}}||_{\infty},||{\mathsf{W}}||_{\infty}\right\}$.}

{\it Proof.} First, we assume that $d_B\leq{d_A}$. Let
${\xx}={\mathsf{Q}}-{\mathsf{R}}$ be the Jordan decomposition of
${\xx}$, then $\Phi({\xx})=\Phi({\mathsf{Q}})-\Phi({\mathsf{R}})$.
It follows from $\Phi({\xx})^{\dagger}=\Phi({\xx})$, Lemma 2.2 and
properties of the trace that
\begin{equation}
||\Phi({\xx})||_{(k)}={\rm{Tr}}_B
\left[\bigl({\mathsf{\Pi}}_{Q}-{\mathsf{\Pi}}_{R}\bigr)
\bigl(\Phi({\mathsf{Q}})-\Phi({\mathsf{R}})\bigr)\right]
\leq{\rm{Tr}}_{B}
\left[\bigl({\mathsf{\Pi}}_{Q}+{\mathsf{\Pi}}_{R}\bigr)
\bigl(\Phi({\mathsf{Q}})+\Phi({\mathsf{R}})\bigr)\right]=\eta{\,}
{\rm{Tr}}_{A}\bigl[({\mathsf{S}}+{\mathsf{T}})|{\xx}|{\,}\bigl]
\label{phiakpl}
\end{equation}
for two mutually orthogonal projectors with
${\rm{rank}}({\mathsf{\Pi}}_{Q}+{\mathsf{\Pi}}_{R})\leq{k}$. In
(\ref{phiakpl}) we use ${\mathsf{Q}}+{\mathsf{R}}=|{\xx}|$ and
positive semidefinite operators
\[
{\mathsf{S}}=\eta^{-1}\sum\nolimits_{n} {\mathsf{E}}_{n}^{\dagger}{\,}{\mathsf{\Pi}}_{Q}{\,}{\mathsf{E}}_{n}
\ , \quad {\mathsf{T}}=\eta^{-1}\sum\nolimits_{n} {\mathsf{E}}_{n}^{\dagger}{\,}{\mathsf{\Pi}}_{R}{\,}{\mathsf{E}}_{n}
\ .
\]
Denoting $\mu\equiv||{\mathsf{M}}||_{\infty}$ and
$\nu\equiv||{\mathsf{W}}||_{\infty}$, we obviously write
$\mu^{-1}{\mathsf{M}}\leq{\mathbf{I}}_B$ and
$\nu^{-1}{\mathsf{W}}\leq{\mathbf{I}}_A$. Combining the former
with properties of the trace, we have
\begin{equation}
{\rm{Tr}}_{A}\bigl({\mathsf{S}}+{\mathsf{T}}\bigr)=\eta^{-1}{\,}{\rm{Tr}}_{B}
\left[\bigl({\mathsf{\Pi}}_{Q}+{\mathsf{\Pi}}_{R}\bigr){\,}{\mathsf{M}}\right]
\leq{\rm{Tr}}_{B}
\left[\bigl({\mathsf{\Pi}}_{Q}+{\mathsf{\Pi}}_{R}\bigr){\,}\mu^{-1}{\mathsf{M}}\right]\leq{k}
\ . \label{trtrmm}
\end{equation}
Using ${\mathsf{\Pi}}_{Q}+{\mathsf{\Pi}}_{R}\leq{\mathbf{I}}_B$
and $\nu^{-1}{\mathsf{W}}\leq{\mathbf{I}}_A$, we also obtain
\begin{equation}
\langle{\uu}{,\,}({\mathsf{S}}+{\mathsf{T}}){\,}{\uu}\rangle=\eta^{-1}\sum\nolimits_{n}
\langle{\uu}{,\,}{\mathsf{E}}_{n}^{\dagger}\bigl({\mathsf{\Pi}}_{Q}+{\mathsf{\Pi}}_{R}\bigr){\mathsf{E}}_{n}{\uu}\rangle
\leq\langle{\uu}{,\,}\eta^{-1}{\mathsf{W}}{\,}{\uu}\rangle
\leq\langle{\uu}{,\,}\nu^{-1}{\mathsf{W}}{\,}{\uu}\rangle\leq\langle{\uu}{,\,}{\uu}\rangle
\label{bcbcww}
\end{equation}
for each ${\uu}\in{\mathscr{H}}_{A}$. This implies
${\mathsf{S}}+{\mathsf{T}}\leq{\mathbf{I}}_A$ and the truth of
using Ky Fan's  principle for the right-hand side of
(\ref{phiakpl}). So, the relations (\ref{phiakpl}) and
(\ref{trtrmm}) provide the claim. When $d_B>d_A$, the
calculations (\ref{bcbcww}) remain valid for $k>d_A$, hence the
right-hand side of (\ref{phiakpl}) is not greater than
$\eta{\,}||{\xx}||_{\rm{tr}}$. $\square$

As it is known, the role of particular symmetric gauge functions
${\rm{g}}_{(k)}(\centerdot)$ is that norm inequalities can
sometimes be extended to all unitarily invariant norms. Let
${\uu},{\vv}\in{\mathbb{C}}^d$ be given vectors with
$d=\max\{d_A,d_B\}$. In accordance with theorem 7.4.45 in
\cite{hornj}, the inequality ${\rm{g}}({\uu})\leq{\rm{g}}({\vv})$
holds for all symmetric gauge functions ${\rm{g}}(\centerdot)$ on
${\mathbb{C}}^d$ if and only if
${\rm{g}}_{(k)}({\uu})\leq{\rm{g}}_{(k)}({\vv})$ for
$k=1,2,\ldots,d$. By Theorem 3.1, for any symmetric gauge function
we then obtain
\[
{\rm{g}}\bigl(\sigma_i(\Phi({\xx}))\bigr)\leq\eta{\ }{\rm{g}}\bigl(\sigma_i({\xx})\bigr)
\ ,
\]
or merely $|||\Phi({\xx})|||_{\rm{g}}\leq \eta{\,}|||{\xx}|||_{\rm{g}}$, whenever ${\xx}\in{\lsa}({\ch}_A)$. In terms of shrinking factors, the norm inequality can
be reformulated as follows.

{\bf Theorem 3.2.} {\it For each unitarily invariant norm
$|||\centerdot|||_{\rm{g}}$, defined on both the spaces ${\ch}_A$
and ${\ch}_B$, a corresponding shrinking factor satisfies}
\[
\eta_{\rm{g}}(\Phi_{s.a.})\leq
\max\left\{||{\mathsf{M}}||_{\infty},||{\mathsf{W}}||_{\infty}\right\}
\ .
\]

In the next section we will show that $||{\mathsf{M}}||_{\infty}$
is the exact value of shrinking factor for the spectral norm and
$||{\mathsf{W}}||_{\infty}$ is the one for the trace norm. So, a
degree of non-contractivity of $\Phi_{s.a.}$ is quite revealed by
these two values.

\section{The spectral norm and trace norm}

Let ${\xx}$ be Hermitian operator such that
$||{\xx}||_{\infty}=1$. Using the Jordan decomposition
${\xx}={\mathsf{Q}}-{\mathsf{R}}$, we get
\begin{equation}
||\Phi({\xx})||_{\infty}=\Bigl|{\rm{Tr}}_B
\bigl[{\piq}\bigl(\Phi({\mathsf{Q}})-\Phi({\mathsf{R}})\bigr)\bigr]\Bigr|\leq
{\rm{Tr}}_B\bigl[{\piq}\bigl(\Phi({\mathsf{Q}})+\Phi({\mathsf{R}})\bigr)\bigr]
\label{espec1}
\end{equation}
for corresponding projector ${\piq}$ of rank one. Due to
$|{\xx}|\leq{\mathbf{I}}_A$ and Ky Fan's maximum principle
(\ref{kfmaxpr02}), the right-hand side of (\ref{espec1}) can be
treated as
\[
{\rm{Tr}}_A\Bigl(\sum\nolimits_{n} {\mathsf{E}}_{n}^{\dagger}{\,}{\piq}{\,}{\mathsf{E}}_{n}{\,}|{\xx}|\Bigr)
\leq{\rm{Tr}}_A\Bigl(\sum\nolimits_{n} {\mathsf{E}}_{n}^{\dagger}{\,}{\piq}{\,}{\mathsf{E}}_{n}\Bigr)
={\rm{Tr}}_B\bigl({\mathsf{M}}{\piq}\bigr)\leq||{\mathsf{M}}||_{\infty}
\ .
\]
So, we have
$||\Phi({\xx})||_{\infty}\leq||{\mathsf{M}}||_{\infty}$ for any
${\xx}\in{\lsa}({\ch}_A)$ with $||{\xx}||_{\infty}=1$. Noting
$\Phi({\mathbf{I}}_A)={\mathsf{M}}$, the inequality between norms
is saturated. Hence we obtain the exact value of shrinking factor
\begin{equation}
\eta_{\infty}(\Phi_{s.a.})=||{\mathsf{M}}||_{\infty}
\ . \label{espec3}
\end{equation}
Note that this is a particular case of the Russo-Dye theorem (see,
e.g., corollary 2.9 in \cite{pauls}). The above calculation is
given here due to its simplicity and illustration of the method.

In line with (\ref{phiakpl}), for the trace norm there holds
\begin{equation}
||\Phi({\xx})||_{\rm{tr}}\leq{\rm{Tr}}_B
\left[\bigl({\piq}_{Q}+{\piq}_{R}\bigr)
\bigl(\Phi({\mathsf{Q}})+\Phi({\mathsf{R}})\bigr)\right]=
{\rm{Tr}}_A\bigl({\mathsf{W}}|{\xx}|\bigr)
\ , \label{etrec1}
\end{equation}
since ${\piq}_{Q}+{\piq}_{R}={\mathbf{I}}_B$ by
${\rm{rank}}({\piq}_{Q}+{\piq}_{R})=d_B$. If
$||{\xx}||_{\rm{tr}}=1$ then the right-hand side of (\ref{etrec1})
does not exceed $||{\mathsf{W}}||_{\infty}$. This value can
actually be reached. Let ${\yy}$ be projector onto 1-dimensional
eigenspace corresponding to the largest eigenvalue of operator
${\mathsf{W}}$. Then $\Phi({\yy})$ is positive semidefinite and
\[
||\Phi({\yy})||_{\rm{tr}}={\rm{Tr}}_B\bigl(\Phi({\yy})\bigr)=
{\rm{Tr}}_A\bigl({\mathsf{W}}{\yy}\bigr)=||{\mathsf{W}}||_{\infty}
\ . \label{etrec2}
\]
In other words, the exact value of shrinking factor is given by
\begin{equation}
\eta_{\rm{tr}}(\Phi_{s.a.})=||{\mathsf{W}}||_{\infty}
\ . \label{etrec3}
\end{equation}
Thus, for the spectral and trace norms the exact value of
shrinking factor is simply calculated. For other norms a task is
more difficult but the bound of Theorem 3.2 is useful for many
aims. So, this bound can be rewritten as
\begin{equation}
\eta_{\rm{g}}(\Phi_{s.a.})\leq
\max\bigl\{\eta_{\infty}(\Phi_{s.a.}),\eta_{\rm{tr}}(\Phi_{s.a.})\bigr\}
\ . \label{etgintr}
\end{equation}
To sum up, we have a valuable conclusion. If the restriction
$\Phi_{s.a.}$ is contractive with respect to both the spectral and
trace norm then it is contractive with respect to all unitarily
invariant norms. Moreover, a degree of non-contractivity can be
measured by using these two norms.

Finally, we apply our results to the operation of partial trace.
This operation is especially important in the context of quantum
information processing. Hence we are interested in relations
between norms before and after partial trace. The writers of
\cite{lidar1} resolved a question for those unitarily invariant
norms that are multiplicative over tensor products. The explicit
Choi-Kraus representation of partial trace is given in
\cite{nielsen}. However, the operators ${\mathsf{M}}$ and
${\mathsf{W}}$ can be found directly. Let us take
${\ch}_A={\ch}_B\otimes{\ch}_C$ with partial tracing over
${\mathscr{H}}_C$, that is
\begin{equation}
\Psi({\xx}):={\rm{Tr}}_C({\xx})
\label{phtrc}
\end{equation}
for any ${\xx}\in{\cal{L}}\bigl({\ch}_B\otimes{\ch}_C\bigr)$.
First, this operation preserves trace, because
\[
{\rm{Tr}}_B\bigl(\Psi({\xx})\bigr)={\rm{Tr}}_B\bigl\{{\rm{Tr}}_C({\xx})\bigr\}
={\rm{Tr}}_A({\xx})
\ .
\]
Combining this with
${\rm{Tr}}_B\bigl(\Psi({\xx})\bigr)={\rm{Tr}}_A\bigl({\mathsf{W}}{\,}{\xx}\bigr)$
finally gives ${\mathsf{W}}={\mathbf{I}}_A$. Second, the
right-hand side of definition for ${\mathsf{M}}$ is rewritten as
\[
\sum\nolimits_{n} {\mathsf{E}}_{n}{\,}{\mathbf{I}}_A{\,}{\mathsf{E}}_{n}^{\dagger}=
\Psi\bigl({\mathbf{I}}_B\otimes{\mathbf{I}}_C\bigr)=
{\mathbf{I}}_B{\>}{\rm{Tr}}_C({\mathbf{I}}_C)
\ .
\]
So we obtain ${\mathsf{M}}=d_C{\,}{\mathbf{I}}_B$, where
$d_C={\rm{dim}}({\ch}_C)$. Because $||{\mathsf{M}}||_{\infty}=d_C$
and $||{\mathsf{W}}||_{\infty}=1$, the statement of Theorem 3.2
gives
\begin{equation}
|||\Psi({\xx})|||_{\rm{g}}\leq d_C{\,}|||{\xx}|||_{\rm{g}}
\label{crlphaa1}
\end{equation}
for ${\xx}\in{\lsa}({\ch}_A)$ and any unitarily invariant norm.
For the spectral norm this relation coincides with the one given
in \cite{lidar1}. For the Frobenius norm the method of
\cite{lidar1} provides a more precise bound. On the other hand,
the validity of (\ref{crlphaa1}) is not restricted to those norms
that are multiplicative over tensor product.


\newpage

\begin{thebibliography}{23}

\bibitem{bhatia07}
Bhatia, R.: {\it Positive Definite Matrices.} Princeton: Princeton University Press, 2007

\bibitem{bhatia87}
Bhatia, R.: Some inequalities for norm ideals. Commun. Math. Phys.
\textbf{111}, 33--39 (1987)

\bibitem{choi}
Choi, M.-D.: Completely positive linear maps on complex matrices.
Linear Algebra Appl. \textbf{10}, 285-–290 (1975)

\bibitem{devetak}
Devetak, I., Junge, M., King, C. and Ruskai, M.B.:
Multiplicativity of completely bounded $p$-norms implies a new
additivity result. Commun. Math. Phys. \textbf{266}, 37--63 (2006)

\bibitem{kyfan}
Fan, K.: On a theorem of Weyl concerning eigenvalues of linear
transformations. I. Proc. Nat. Acad. Sci. USA \textbf{35},
652--655 (1949)

\bibitem{hornj}
Horn, R.A. and Johnson, C.R.: {\it Matrix Analysis.} Cambridge: Cambridge University Press, 1985

\bibitem{jenca}
Jen\v{c}ov\'{a}, A.: A relation between completely bounded norms
and conjugate channels. Commun. Math. Phys. \textbf{266}, 65–-70
(2006)

\bibitem{kraus}
Kraus, K.: General state changes in quantum theory. Ann. Phys. \textbf{64}, 311-–335 (1971)

\bibitem{lidar1}
Lidar, D.A., Zanardi, P. and Khodjasteh, K.: Distance bounds on
quantum dynamics. Phys. Rev. A \textbf{78}, 012308 (2008)

\bibitem{nielsen}
Nielsen, M.A. and Chuang, I.L.: {\it Quantum Computation and
Quantum Information.} Cambridge: Cambridge University Press, 2000

\bibitem{pauls}
Paulsen, V.: {\it Completely Bounded Maps and Operator Algebras.}
Cambridge: Cambridge University Press, 2002

\bibitem{ruskai}
Peres-Garsia, D., Wolf, M.M., Petz, D. and Ruskai, M.B.:
Contractivity of positive and trace preserving maps under $L_p$
norms. J. Math. Phys. \textbf{47}, 083506 (2006)

\bibitem{rast07}
Rastegin, A.E.: Trace distance from the viewpoint of quantum
operation techniques. J. Phys. A: Math. Theor. \textbf{40},
9533--9549 (2007)

\bibitem{rast092}
Rastegin, A.E.: Some properties of partial fidelities. Quantum
Inf. Comput. \textbf{9}, 1069--1080 (2009)

\bibitem{rast091}
Rastegin, A.E.: Partitioned trace distances. Quantum Inf. Process.
DOI 10.1007/s11128-009-0128-7

\bibitem{uhlmann77}
Uhlmann, A.: Relative entropy and the Wigner-Yanase-Dyson-Lieb
concavity in an interpolation theory. Commun. Math. Phys.
\textbf{54}, 21--32 (1977)

\bibitem{uhlmann00}
Uhlmann, A.: On ''partial'' fidelities. Rep. Math. Phys. \textbf{45}, 407--418 (2000)

\bibitem{watrous2}
Watrous, J.: Notes on super-operator norms induced by Schatten
norms. Quantum Inf. Comput. \textbf{5}, 58–-68 (2005)

\end{thebibliography}
\end{document}